\documentclass[floats,floatfix,showpacs,amssymb,prd,superscriptaddress,twocolumn,aps,nofootinbib]{revtex4-1}
\usepackage{graphicx, epsfig,amssymb} %include figure files
\usepackage{epstopdf}
\usepackage{amsmath, amsfonts}
\usepackage{bm} %include bold math: \bm{} creates bold letters in math mode
\usepackage[breaklinks]{hyperref}
\usepackage{color}

\def\be{\begin{equation}}
\def\ee{\end{equation}}

\begin{document}

\title{Tidal effects around higher-dimensional black holes}

\author{Richard Brito}
\email{richard.brito@ist.utl.pt}
\affiliation{CENTRA, Departamento de F\'{\i}sica, Instituto Superior T\'ecnico, Universidade T\'ecnica de Lisboa - UTL,
Avenida~Rovisco Pais 1, 1049 Lisboa, Portugal.}

\author{Vitor Cardoso}\email{vitor.cardoso@ist.utl.pt}
\affiliation{CENTRA, Departamento de F\'{\i}sica, Instituto Superior T\'ecnico, Universidade T\'ecnica de Lisboa - UTL,
Avenida~Rovisco Pais 1, 1049 Lisboa, Portugal.}
\affiliation{Department of Physics and Astronomy, The University of Mississippi, University, MS 38677, USA.}

\author{Paolo Pani}\email{paolo.pani@ist.utl.pt}
\affiliation{CENTRA, Departamento de F\'{\i}sica, Instituto Superior T\'ecnico, Universidade T\'ecnica de Lisboa - UTL,
Avenida~Rovisco Pais 1, 1049 Lisboa, Portugal.}

\date{\today}

\begin{abstract}
In four-dimensional spacetime, moons around black holes generate low-amplitude tides, and the energy extracted from the hole's rotation
is always smaller than the gravitational radiation lost to infinity. Thus, moons orbiting a black hole inspiral and eventually merge.
However, it has been conjectured that in higher-dimensional spacetimes orbiting bodies generate much stronger tides, which backreact by tidally accelerating the body {\it outwards}. This effect, analogous to the tidal acceleration experienced by the Earth-Moon system, would determine the evolution of the binary. Here, we put this conjecture to the test, by studying matter coupled to a massless scalar field in orbit around a singly-spinning rotating black hole in higher dimensions. 
We show that in dimensions larger than five the energy extracted from the black hole through superradiance is larger than the energy carried out to infinity.
Our numerical results are in excellent agreement with analytic approximations and lend strong support to the conjecture that tidal acceleration is the rule, rather than the exception, in higher dimensions. Superradiance dominates the energy budget and moons ``outspiral''; for some particular orbital frequency, the energy extracted at the horizon equals the energy emitted to infinity and ``floating orbits'' generically occur. We give an interpretation of this phenomenon in terms of the membrane paradigm and of tidal acceleration due to energy dissipation across the horizon.  
\end{abstract}


%04.70.-s 	Physics of black holes (see also 97.60.Lf—in astronomy)
%02.60.Lj %Ordinary and partial differential equations; boundary value problems
%04.25.-g 	Approximation methods; equations of motion
%04.50.Gh Higher-dimensional black holes, black strings, and related objects 
%96.15.Wx 	Tidal forces 
\maketitle
%%%%%%%%%%%%%%%%%%%%%%%%%%%%%%%%%%%%%%%%%%%%%%%%%%%%%%%%
\section{Introduction}
%%%%%%%%%%%%%%%%%%%%%%%%%%%%%%%%%%%%%%%%%%%%%%%%%%%%%%%%

Gravitational binaries are intrinsically complicated systems which display a wealth of interesting effects. One important effect, which occurs for example in several planet-moon systems, are tides generated by differential gravitational forces.
Tidal forces in the Earth-Moon system have long ago locked the Moon in a synchronous rotation with the Earth and have increased the Earth-Moon distance~\cite{darwin,hut}.
These processes, so-called tidal locking and tidal acceleration, respectively, are only possible if there is some dissipation mechanism in the system. Because  of friction, tides extract energy from the binary and, since angular momentum must be conserved, this provides a mechanism to exchange angular momentum between the Earth and the Moon.
In the Earth-Moon case the dissipation is caused by the friction between the oceans and the Earth surface. In binaries containing rotating black holes (BHs) the event horizon can play the role of a dissipative membrane and tidal acceleration is known for many years under a different name: \emph{superradiance}~\cite{zeldo1,zeldo2,teunature,Cardoso:2012zn}. 

Consider a Kerr BH whose angular velocity of the horizon is $\Omega_H$. A wave with frequency 
$\omega<m\Omega_H$ scattering off the BH (where m is the azimuthal quantum number) is amplified, and it extracts rotational energy from the BH.
Superradiance is responsible for many interesting effects (e.g.~\cite{misner,teunature,Cardoso:2004nk,Cardoso:2007az,Cardoso:2011xi,Cardoso:2012zn,Yoshino:2012kn}), one of them is the possible existence of ``floating orbits'' around BHs. Generically, orbiting bodies around BHs spiral {\it inwards} as a consequence of gravitational-wave emission. 
When the condition for superradiance is met, it is possible to imagine the existence of floating orbits, i.e., orbits in which the energy radiated to infinity by the body is entirely compensated by the energy extracted from the BH~\cite{misner,teunature}. Within general relativity in four dimensions, tidal effects are in general completely washed-out by gravitational-wave emission and orbiting bodies always spiral inwards \cite{Cardoso:2012zn}. However, when coupling to scalar fields is allowed, an induced dipole moment produces a tidal acceleration (or polarization acceleration \cite{Cardoso:2012zn}) which might be orders of magnitude stronger than tidal quadrupolar effects.
Furthermore, in theories where massive scalar fields are present, the coupling of the scalar field to matter can produce resonances in the scalar energy flux, which can lead to floating orbits outside the innermost stable circular orbit~\cite{Cardoso:2011xi,Yunes:2011aa}.

It was recently argued via a tidal analysis framework that higher-dimensional BHs in general relativity should be prone to strong tidal effects \cite{Cardoso:2012zn}. One of the consequences of those studies was that orbiting bodies around higher-dimensional rotating BHs always spiral \emph{outwards}, if the tidal acceleration (or, equivalently, the superradiance) condition is met. In this paper, we use a fully relativistic analysis, albeit in the test-particle limit, to prove this behavior.
For simplicity, we consider the coupling of massless scalar fields to matter around a rotating BH in higher-dimensional spacetimes. We show that, for spacetime dimensions
$D>5$, tidal effects are so strong, that the energy extracted from the BH is greater than the energy radiated to infinity. Higher-dimensional spacetimes are of interest in a number of theories and scenarios~\cite{Cardoso:2012qm}, in our case we view them as a proof-of-principle for strong tidal effects in BH physics, without the need for resonances.
We do not consider gravitational perturbations, gravitational effects should be subdominant with respect to the dipolar effects discussed here \cite{Cardoso:2012zn}. Nevertheless, the arguments
presented in Ref.~\cite{Cardoso:2012zn} together with the present results show that a purely gravitational interaction also displays this phenomenon, which likely leads to new interesting effects
in higher dimensional BH physics.

This paper is organized as follows. In Section~\ref{sec:tides}, extending the discussion of Ref.~\cite{Cardoso:2012zn}, we study the effect of an electrically charged particle orbiting a neutral central object in $D=4+n$ dimensions. By applying the membrane paradigm~\cite{Thorne:1986iy}, we derive a simple formula for the ratio between the energy flux at infinity and at the horizon in $4+n$ dimensions.  In Section~\ref{sec:myers} we compute the energy fluxes in terms of BH perturbations sourced by a test-particle in circular orbit around a spinning BH. We derive the Teukolsky equations and the expressions for the energy fluxes. 
In Section~\ref{sec:analytical} we solve the wave equation analytically in the low-frequency regime. We then compare our analytical results with those obtained by a direct numerical integration of the wave equation, as discussed in Section~\ref{sec:numerical}.
We conclude in Section~\ref{sec:conclusion}. Throughout the paper we use $G=c=1$ units, except in Sec.~\ref{sec:tides} where, for clarity, we show $G$ and $c$ explicitly.    
%%%%%%%%%%%%%%%%%%%%%%%%%%%%%%%%%%%%%%%%%%%%%%%%%%%%%%%%%%%%%%%%%%%%%%%%%%%%%%
\section{Tides for charged interactions in $4+n$ dimensions}\label{sec:tides}
%%%%%%%%%%%%%%%%%%%%%%%%%%%%%%%%%%%%%%%%%%%%%%%%%%%%%%%%%%%%%%%%%%%%%%%%%%%%%%
The flux emitted by a particle orbiting a spinning BH can be estimated at newtonian level in terms of BH tidal acceleration and by applying the membrane paradigm~\cite{Cardoso:2012zn}. In this section, we generalize the computation sketched in Ref.~\cite{Cardoso:2012zn} to higher dimensions and to scalar fields.

Let us consider the interaction of a particle with scalar charge $q_p$ and gravitational mass $m_p$ orbiting a neutral central object of mass $M$ and radius $R$. If the object has a dielectric constant $\epsilon=\epsilon_r\epsilon_0$, the particle external field induces a polarization surface charge density on the central object and a dipole moment which are given respectively by~\cite{jackson}
\begin{align}
\sigma_{\rm pol}&=(3+n)\epsilon_0\beta E_0 \cos\vartheta\,,\\
p&=\Omega_{(n+3)}\epsilon_0\beta R^{3+n}E_0\,, \label{dipmom}
\end{align}
where 
\begin{equation}
 E_0= \frac{q_p}{\Omega_{(n+3)}\epsilon_0 r_0^{2+n}}\,,
\end{equation}
and $r_0$ is the orbital distance, $\Omega_{(n+3)}$ is the solid angle of the $(n+3)$-sphere, $\beta$ is some constant that depends on the relative dielectric constant of the object and $\vartheta$ is the polar angle with respect to single axis of rotation of the central object. 

Assuming circular orbits, the tangential force on the charge $q_p$ due to the induced electric field is given by
\be 
F_{\vartheta}=\frac{q_p p}{\Omega_{(n+3)} \epsilon_0 r_0^{3+n}}\sin\vartheta\,.
\ee
Without dissipation, the dipole moment would be aligned with the particle's position vector. Here, we consider that dissipation introduces a small time lag $\tau$, such that the dipole moment leads the particle's position vector by a constant angle $\phi$ given by (see~\cite{hut,Cardoso:2012zn} for details)
\be 
\phi=(\Omega-\Omega_H)\tau\,,
\ee
where $\Omega_H$ and $\Omega$ are the rotational angular velocity and the orbital angular velocity, respectively.
At first order in $\phi$, the tangential component of the force reads
\be 
F_{\vartheta}\sim\frac{q_p p}{\Omega_{(n+3)} \epsilon_0 r_0^{3+n}}(\Omega-\Omega_H)\tau\,.
\ee
This exerts a torque $r_0 F_{\vartheta}$ and the change in orbital energy over one orbit reads
\begin{eqnarray}
\dot E_{\rm orbital}&=&\frac{1}{2\pi}\int_0^{2\pi} r_0 F_{\vartheta}{\Omega}\,d\vartheta=\Omega r_0 F_{\vartheta}\nonumber\\
&=&\frac{\beta q^2_p\,R^{3+n}}{\Omega_{(n+3)} \epsilon_0 r_0^{4+2n}}\Omega(\Omega-\Omega_H)\tau\,,
\end{eqnarray}
where, in the last step, we used Eq.~\eqref{dipmom}.

Remarkably, the equation above qualitatively describes the energy flux across the horizon of a rotating BH if one identifies $\Omega_H$ with the angular velocity of the BH and the lag $\tau$ with the light-crossing time, $\tau\sim R/c$, where $R^{1+n}=G_D M/[(n+1)c^2]$, $G_D$ is the $D$-dimensional gravitational constant and $M$ is the BH mass~\cite{Cardoso:2012zn}. Accordingly, a particle orbiting a rotating BH in $4+n$ dimensions dissipates energy at the event horizon at a rate of roughly
\be 
\dot E_{H}=\frac{\beta\,q^2_p\,G_D^{\frac{4+n}{1+n}}}{\Omega_{(n+3)} \epsilon_0 c^{\frac{3(n+3)}{1+n}}(n+1)^{\frac{4+n}{1+n}}}
\frac{M^{\frac{4+n}{1+n}}}{r_0^{4+2n}}\Omega(\Omega-\Omega_H)\,.\label{EdotH}
\ee
On the other hand, charged accelerating particles radiate to infinity according to Larmor's formula, which in $4+n$ dimensions reads~\cite{Cardoso:2007uy} (this will be also derived below, see Eq.~\eqref{anafluxinf})
\be 
\dot E_{\infty}=\frac{\gamma^2}{c^{3+n}}\Omega^{4+n}r_0^2\,,\label{EdotINF}
\ee
where $\gamma$ is some coupling constant.

Tidal acceleration~\cite{Cardoso:2012zn} occurs when the orbit of the particle is pushed outwards due to energy dissipation in the central object.
This is only possible if two conditions are satisfied: (i) $\Omega<\Omega_H$, so that $\dot{E}_H<0$ and the energy flows out of the BH; and (ii) $|\dot E_H|>\dot E_{\infty}$, i.e. the rate at which energy is dissipated to infinity must be smaller than the rate at which energy is extracted from the BH. From Eqs.~\eqref{EdotH} and~\eqref{EdotINF}, and using $\Omega\sim r_0^{-(3+n)/2}$~\cite{Cardoso:2008bp}, we find
\begin{align}\label{tideratio}
\frac{|\dot E_H|}{\dot E_{\infty}}&=\frac{\beta q_p^2 G_D^{\frac{4+n}{1+n}}}{\Omega_{(n+3)} \epsilon_0 \gamma^2 c^{\frac{(n+3)(2-n)}{n+1}}(n+1)^{\frac{4+n}{1+n}}}\times\nonumber\\
&\frac{M^{\frac{4+n}{1+n}}}{\Omega^{3+n}r_0^{2(3+n)}}(\Omega_H-\Omega)
\sim \left(\frac{v}{c}\right)^{-\frac{(n-1)(n+3)}{n+1}}\,.
\end{align}
where we have assumed $\Omega_H\gg\Omega$ and we have defined the orbital velocity 
%%%
\begin{equation}
 v=\left[M(n+1)\right]^{\frac{1}{n+3}}\Omega^{\frac{n+1}{n+3}}\,.\label{vel}
\end{equation}

%%%
At large distance, $v\sim r_0\Omega$ and, when $n=0$, we recover the standard definition, $v=(M\Omega)^{1/3}$.
Surprisingly, for $n>1$ ($D>5$) tidal acceleration dominates at large distances. This simple argument suggests that test-particles orbiting rotating BHs in dimensions greater than five would generically extract energy from the BH horizon at a larger rate than the energy emitted in gravitational waves to infinity. As a consequence, the orbital separation will increase in time, i.e. the system will ``outspiral''. In the next sections we shall prove this is indeed the case, by computing the linear response of a higher-dimensional spinning BH to a test-particle in circular orbit.
%%%%%%%%%%%%%%%%%%%%%%%%%%%%%%%%%%%%%%%%%%%%%%%%%%%%%%%%%%%%%%%%%%%%%%%%%%%%%%%%%%%%%%%%%%%%%%%
\section{Scalar perturbations of singly-spinning Myers-Perry black holes}\label{sec:myers}
%%%%%%%%%%%%%%%%%%%%%%%%%%%%%%%%%%%%%%%%%%%%%%%%%%%%%%%%%%%%%%%%%%%%%%%%%%%%%%%%%%%%%%%%%%%%%%%
\subsection{The background metric}
%%%%%%%%%%%%%%%%%%%%%%%%%%%%%%%%%%%%%%%%%%%%%%%%%%%%%%%%%%%%%%%%%%%%%%%%%%%%%%%%%%%%%%%%%%%%%%%
In four dimensions, there is only one possible angular momentum parameter for an axisymmetric spacetime and rotating BH solutions are uniquely described by the Kerr family. In higher dimensions there are several choices of rotation axis, which correspond to a multitude of angular momentum parameters~\cite{myers}. Here we shall focus on the simplest case, where there is only a single axis of rotation. In the following we shall adopt the notation used in Refs.~\cite{Ida,Cardoso:2004cj,Cardoso:2005vk}, to which we refer for details.
 
The metric of a $4 + n$ dimensional Kerr-Myers-Perry BH with only one nonzero angular momentum parameter is given in Boyer-Lindquist coordinates by~\cite{myers}
\begin{align} 
\label{Myers-Perry}
ds^2&=-\frac{\Delta-a^2\sin^2\vartheta}{\Sigma}dt^2-\frac{2a(r^2+a^2-\Delta)\sin^2\vartheta}{\Sigma}dtd\phi\nonumber\\
&+\frac{(r^2+a^2)^2-\Delta a^2\sin^2\vartheta}{\Sigma}\sin^2\vartheta d\phi^2+\frac{\Sigma}{\Delta}dr^2\nonumber\\
&+\Sigma d\vartheta^2+r^2\cos^2\vartheta d\Omega_n^2\,,
\end{align}
where
\begin{equation}
\Sigma= r^2+a^2\cos^2\vartheta\,,\qquad \Delta= r^2+a^2-2M r^{1-n}\,,
\end{equation}
and $d\Omega_n^2$ denotes the standard line element of the unit $n$-sphere. 
This metric describes a rotating BH in asymptotically flat, vacuum spacetime, whose physical mass ${\cal M}$ and angular momentum ${\cal J}$ (transverse to the $r\phi$ plane) respectively read
%%%
\begin{equation}
 {\cal M}=\frac{(n+2) A_{n+2}}{8\pi}M\,,\qquad {\cal J}=\frac{2}{n+2} {\cal M} a\,,
\end{equation}
%%%
where $A_{n+2}=2\pi^{{n+3}/2}/\Gamma[(n+3)/2]$.% and we assume $\mu, a>0$.

The event horizon is located at $r=r_H$, defined as the largest real root of $\Delta$. In four dimensions, an event horizon exists only for $a\leq M$. In five dimensions, an event horizon exists only for $a\leq\sqrt{2M}$, and the BH area shrinks to zero in the extremal limit $a\rightarrow \sqrt{2M}$. On the other hand, when $D>5$, there is no upper bound on the BH spin and a horizon exists for any $a$.
%%%%%%%%%%%%%%%%%%%%%%%%%%%%%%%%%%%%%%%%%%%%%%%%%%%%%%%%
\subsection{Setup}
%%%%%%%%%%%%%%%%%%%%%%%%%%%%%%%%%%%%%%%%%%%%%%%%%%%%%%%%
We consider a small object orbiting a spinning BH and a massless scalar field coupled to matter. At first order in perturbation theory, the scalar field equation in the background~(\ref{Myers-Perry}) reads
\be\label{fieldeq}
\square\varphi\equiv\frac{1}{\sqrt{-g}}\frac{\partial}{\partial x^{\mu}}\left(\sqrt{-g}g^{\mu\nu}\frac{\partial}{\partial x^{\nu}}\varphi\right)
=\alpha {\cal T}\,,
\ee
where $\alpha$ is some coupling constant. For simplicity we focus on source terms of the form
\be
{\cal T}=\int \frac{d\tau}{\sqrt{-g}} m_p\delta^{(4+n)}(x-X(\tau))\,,
\ee
which corresponds to the trace of the stress-energy tensor of a point particle with mass $m_p$. We also restrict to equatorial circular orbits ($\dot\vartheta=0$, $\vartheta=\pi/2$), which is an unrealistic approximation in higher dimensions: generic circular orbits
are unstable, with an instability timescale of order of the orbital period \cite{Cardoso:2008bp}. Nevertheless, our purpose here is to show that tidal effects can dominate, it is not clear what the overall combined effect of tidal acceleration and circular geodesic motion instability is. Extending the present analysis to generic orbits and relaxing the test-particle approximation are interesting future developments.

For prograde orbits around a singly-spinning Myers-Perry BH~(\ref{Myers-Perry}) the energy, angular momentum and frequency of the point particle with mass $m_p$ orbiting at $r=r_0$ read~\cite{Cardoso:2008bp}
\be
\frac{E_p}{m_p}=\frac{a \sqrt{(n+1)M}+r_0^{\frac{3+n}{2}}-2Mr_0^{\frac{1-n}{2}}}{r_0^{\frac{3+n}{4}} \sqrt{2 a \sqrt{(n+1)M}+r_0^{\frac{3+n}{2}}-(n+3)Mr_0^{\frac{1-n}{2}}}}\,,\label{Ep}
\ee
\be
\frac{L_p}{m_p}=\frac{\sqrt{(n+1)M} \left(r_0^2-2 a \sqrt{\frac{M}{n+1}}r_0^\frac{1-n}{2}+a^2\right)}{r_0^{\frac{3(n+1)}{4}} \sqrt{2 a \sqrt{(n+1)M}+r_0^{\frac{3+n}{2}}-(n+3)Mr_0^{\frac{1-n}{2}}}}\,,\label{Lp}
\ee
\be
\Omega_p=\frac{\sqrt{(n+1)M}}{a\sqrt{(n+1)M}+r_0^{\frac{3+n}{2}}}\,.\label{Omegap}
\ee
The only nonvanishing components of the $(4+n)$-velocity $U^\nu$ of the particle on a timelike geodesic are given by
%%%
\begin{eqnarray}
 m_p\Delta_{r=r_0} U^t&=& \left(r_0^2+a^2+\frac{2Ma^2}{r_0^{n+1}}\right)E_p-\frac{2Ma L_p}{r_0^{n+1}} \,,\\
 m_p\Delta_{r=r_0}  U^\phi&=& \frac{2Ma E_p}{r_0^{n+1}}+\left(1-\frac{2M}{r_0^{n+1}}\right)L_p
\end{eqnarray}

%%%%%%%%%%%%%%%%%%%%%%%%%%%%%%%%%%%%%%%%%%%%%%%%%%%%%%%%
\subsection{The wave equation}
%%%%%%%%%%%%%%%%%%%%%%%%%%%%%%%%%%%%%%%%%%%%%%%%%%%%%%%%
Because of the coupling to matter, the orbiting object emits scalar radiation which is governed by Eq.~(\ref{fieldeq}). 
To separate Eq.~\eqref{fieldeq}, we consider the ansatz
\be
\varphi(t,r,\vartheta,\phi)=\sum_{l,m,j}\int d\omega e^{im\phi-i\omega t}R(r)S_{lmj}(\vartheta)Y_j\,,
\ee
where $Y_j$ are hyperspherical harmonics~\cite{Berti:2005gp,Cardoso:2004cj} on the $n$-sphere with eigenvalues given by $-j(j+n-1)$ and $j$ being a non-negative integer. The radial and angular equations read
\begin{eqnarray}\label{teuradial}
&&r^{-n}\frac{d}{dr}\left(r^n\Delta\frac{dR}{dr}\right)+\left\{\frac{\left[\omega(r^2+a^2)-ma\right]^2}{\Delta}\right.
\nonumber\\
&&\left.-\frac{j(j+n-1)a^2}{r^2}-\lambda\right\}R=T_{{lmj}}\,,
\end{eqnarray}
and
\begin{align}\label{ang}
\frac{1}{\sin\vartheta\cos^n\vartheta}\frac{d}{d\vartheta}\left(\sin\vartheta\cos^n\vartheta\frac{dS_{lmj}}{d\vartheta}\right)
+\left[\omega^2a^2\cos^2\vartheta \right.\nonumber\\
\left.-\frac{m^2}{\sin^2\vartheta}-
\frac{j(j+n-1)}{\cos^2\vartheta}+A_{lmj}\right]S_{lmj}=0\,,
\end{align}
where $\lambda=A_{lmj}-2m\omega a+\omega^2a^2$ and
\begin{align}
T_{{lmj}}=&-\frac{m_p\alpha}{U^tr^n}S^*_{lmj}(\pi/2)Y_{j}^{*}(\pi/2,\pi/2,\ldots) \nonumber\\
&\times\delta(r-r_0)\delta(m\Omega_p-\omega)\,,
\label{tlmw}
\end{align}
which has been derived from the stress-energy tensor of the point particle.
Defining a new radial function $X_{lmj}(r)$ 
\be
X_{lmj}=r^{n/2}(r^2+a^2)^{1/2}R\,,
\ee
we get the non-homogeneous equation for the scalar field
\be\label{radial}
\left[\frac{d^2}{dr_*^2}+V\right]X_{{lmj}}(r^*)=\frac{\Delta}{(r^2+a^2)^{3/2}}r^{n/2}T_{{lmj}}\,,
\ee
where $dr/dr_*=\Delta/(r^2+a^2)$ defines the standard tortoise coordinates and the effective potential $V$ reads

\begin{eqnarray}
V&&=\omega^2+\frac{3r^2\Delta^2}{(r^2+a^2)^4}-\frac{\Delta\left[3r^2+a^2-2M r^{1-n}(2-n)\right]}{(r^2+a^2)^3}\nonumber\\
&&+\frac{1}{(r^2+a^2)^{2}}\left\{a^2m^2-4M+\frac{am\omega}{r^{n-1}}-\Delta\left(\omega^2a^2+A_{lmj}\right)\right.\nonumber\\
&&+\left.
\Delta\left[\frac{n(2-n)\Delta}{4r^2}-n+\frac{2n\left(1-n\right)M}{2r^{n+1}}\right.\right.\nonumber\\
&&\left.\left.-\frac{j(j+n-1)a^2}{r^{2}}\right]\right\}.\nonumber
\end{eqnarray}

In the low frequency limit the angular equation~(\ref{ang}) can be solved exactly. At first order in $a\omega$, the eigenvalues can be computed analytically~\cite{Berti:2005gp}
\be 
A_{kjm}=(2k+j+|m|)(2k+j+|m|+n+1)+{\cal O}(a\omega)\,.\label{Akjm}
\ee
By setting $2k=l-(j+|m|)$, the eigenvalues above take the form $A_{ljm}=l(l+n+1)$ and $l$ is such that $l\geq(j+|m|)$, which generalizes the four-dimensional case. An important difference from the four-dimensional case is that regularity of the angular eigenfunctions requires $k$ to be a non-negative integer, i.e. for given $j$ and $m$ only specific values of $l$ are admissible. In fact, it is convenient to label the eigenfunctions and the eigenvalues with the ``quantum numbers'' $(k,j,m)$ rather than with $(l,j,m)$ as in the four dimensional case.
The (non-normalized) 0th-order eigenfunctions are given in terms of hypergeometric functions~\cite{Cardoso:2004cj,Berti:2005gp}
\begin{equation} 
S_{kjm}\propto\sin(\vartheta)^{|m|}x^{j}F\Big[-k,k+j+|m|+\frac{n+1}{2},j+\frac{n+1}{2};x^2\Big]\,,\nonumber
\end{equation}
%%%
where $x=\cos(\vartheta)$.
We adopt the following normalization condition
\be
\int_0^{\pi/2} d\vartheta\sin\vartheta\cos^n\vartheta S_{kjm}S^*_{kjm}=1\,,
\ee
where the integration domain has been chosen in order to have a nonvanishing measure also in the case of odd dimensions. Note that this normalization differs from that adopted in Ref.~\cite{Berti:2005gp}.

We note that at $\vartheta=\pi/2$ only hyperspherical harmonics with $j=0$ are non-vanishing. Thus, in order to calculate the fluxes on circular orbits, one only needs to consider terms with $j=0$. In this case, the hyperspherical harmonics $Y_0$ are constant. 

%
%%%%%%%%%%%%%%%%%%%%%%%%%%%%%%%%%%%%%%%%%%%%%%%%%%%%%%%
\subsection{Green function approach and energy fluxes}
%%%%%%%%%%%%%%%%%%%%%%%%%%%%%%%%%%%%%%%%%%%%%%%%%%%%%%%
To solve the wave equation, let us choose two independent solutions $X_{{kjm}}^{r_H}$ and $X_{{kjm}}^{\infty}$ of the homogeneous equation which satisfy the following boundary conditions
\begin{align} \label{bound}
X_{{kjm}}^{\infty}&\sim e^{ik_{\infty} r_*},\nonumber\\
X_{{kjm}}^{r_H}&\sim A_{\rm{out}}e^{ik_{\infty} r_*}+A_{\rm{in}}e^{-ik_{\infty} r_*}, ~~\rm{as} ~~~r\to \infty\,,\nonumber\\
X_{{kjm}}^{\infty}&\sim B_{\rm{out}}e^{ik_H r_*}+B_{\rm{in}} e^{-ik_H r_*}\,,\nonumber\\
X_{{kjm}}^{r_H}&\sim e^{-ik_H r_*}, ~~\rm{as} ~~~r\to r_H\,.
\end{align}
Here $k_H=\omega-m\Omega_H, k_{\infty}=\omega$ and $\Omega_H\equiv-
\lim_{r\to r_H}g_{t\phi}/g_{\phi\phi}={a}/({r_H^2+a^2})$ is the angular velocity at the horizon of locally nonrotating observers.
The Wronskian of the two linearly independent solutions reads
\begin{equation} \label{wronskian}
W=X_{{kjm}}^{r_H}\frac{dX_{{kjm}}^{\infty}}{dr_*}-X_{{kjm}}^{\infty}\frac{dX_{{kjm}}^{r_H}}{dr_*}=2ik_{\infty} A_{\rm in}\,,
\end{equation}
and it is constant by virtue of the field equations.
Finally, Eq.~(\ref{radial}) can be solved in terms of the Green function~\cite{Detweiler:1978ge}
\begin{align}
&X_{{kjm}}(r_*)=\frac{X_{{kjm}}^{\infty}}{W}\int_{-\infty}^{r_*}{{T}}_{{kjm}}(r^{\prime})\frac{\Delta r^{\prime n/2}}{(r^{\prime 2}+a^2)^{3/2}}X_{{kjm}}^{r_H}dr_*^{\prime}\nonumber\\
&+\frac{X_{{kjm}}^{r_H}}{W}\int_{r_*}^{\infty}{{T}}_{{kjm}}(r^{\prime})\frac{\Delta r^{\prime n/2}}{(r^{\prime 2}+a^2)^{3/2}}X_{{kjm}}^{\infty}dr_*^{\prime}\,.
\end{align}
For very large values of $r$ the above equation has the following asymptotic form
\begin{align}
&X_{{kjm}}(r\to\infty)=\nonumber\\
&\frac{e^{i k_{\infty} r_*}}{2ik_{\infty} A_{{\rm{in}}}} \int_{-\infty}^{\infty} T_{{kjm}}(r^{\prime})X_{{kjm}}^{r_H}\frac{\Delta r^{\prime n/2}}{(r^{\prime 2}+a^2)^{3/2}}dr_*^{\prime}\nonumber\\
&=Z_{{kjm}}^{\infty}\delta(\omega-m\Omega_p) e^{i k_{\infty} r_*}\,,\label{rlm}
\end{align} 
where, using Eq.~\eqref{tlmw},  
\be
Z_{{kjm}}^{\infty}=-\alpha\frac{X_{{kjm}}^{r_H}(r_0)}{WU^t}
\frac{S_{kjm}^{*}(\pi/2)Y_{j}^{*}(\pi/2,\pi/2,\ldots)}{\sqrt{r_0^2+a^2}r_0^{n/2}}m_p\,.\label{insol}
\ee
Likewise, at the horizon we get
\be
X_{{kjm}}(r_*\to -\infty)=Z_{{kjm}}^{r_H}\delta(\omega-m\Omega_p)e^{-ik_H r_*}\label{rhor}
\ee
where, 
\be\label{horsol}
Z_{{kjm}}^{r_H}=-\alpha\frac{X_{{kjm}}^{\infty}(r_0)}{W U^t}\frac{S_{kjm}^{*}(\pi/2)Y_{j}^{*}(\pi/2,\pi/2,\ldots)}{\sqrt{r_0^2+a^2}r_0^{n/2}}m_p\,.
\ee
%
%%%%%%%%%%%%%%%%%%%%%%%%%%%%%%%%%%%%%%%%%%%%%%%%%%%%%%%%%%%%%%%%%%%%%%%%%%%
%\subsection{The Energy Flux}
%%%%%%%%%%%%%%%%%%%%%%%%%%%%%%%%%%%%%%%%%%%%%%%%%%%%%%%%%%%%%%%%%%%%%%%%%%%

The scalar energy flux at the horizon and at infinity are defined as
\be
\dot{E}_{H,\infty}=\lim_{r\to r_H,\infty} \int d\vartheta d\phi \prod_{i=1}^n d\vartheta_i \sqrt{-g}T^{r}_{t}\,,
\ee
where the stress tensor reads
\be
T_{\mu\nu}=(\nabla_{\mu}\varphi\nabla_{\nu}\varphi^*-\frac{1}{2}g_{\mu\nu}\nabla_{\alpha}\varphi\nabla^{\alpha}\varphi^*)\,.
\ee
Finally, using Eqs.~\eqref{rlm} and \eqref{rhor}, we get
%%%%%
\be\label{flux}
\dot{E}_{H,\infty}=\sum_{kjm}m\Omega_p k_{H,\infty}|Z^{r_H,\infty}_{{kjm}}|^2 \,.
\ee
%%%%%
The equation above shows that, if the superradiant condition $k_H<0$ ($\omega<m\Omega_H$) is met, the energy flux at the horizon can be \emph{negative}, $\dot{E}_{H}<0$, i.e. energy can be extracted from a spinning BH~\cite{Teukolsky:1974yv,teunature}. In four dimensions, $|\dot{E}_H|\ll\dot{E}_\infty$ and the superradiant extraction is generically negligible. As we show in the next section, in higher dimensions the opposite is true, $|\dot{E}_H|\gg\dot{E}_\infty$ and superradiance dominates over gravitational-wave emission.

%%%%%%%%%%%%%%%%%%%%%%%%%%%%%%%%%%%%%%%%%%%%%%%%%%%%%%%%%%%%%%%%%%%%%%%%%%%%%
\section{Analytical solution at low frequencies}\label{sec:analytical}
%%%%%%%%%%%%%%%%%%%%%%%%%%%%%%%%%%%%%%%%%%%%%%%%%%%%%%%%%%%%%%%%%%%%%%%%%%%%%
The scalar flux can be evaluated analytically in the low-frequency regime (see e.g.~\cite{poisson1,poisson2,kanti,chen,Creek:2007plb}). 
Let us first focus on the solution $X^{r_H}_{{kjm}}$, which is regular at the horizon.

We first make the following change of variable 
\be
h=\frac{\Delta}{r^2+a^2} \Rightarrow \frac{dh}{dr}=(1-h)r\frac{A(r)}{r^2+a^2}\,, 
\ee
where $A(r)=(n+1)+(n-1)a^2/r^2$. Then, near the horizon $r\sim r_H$, the radial equation~(\ref{teuradial}) can be written as
\begin{align}\label{NHrot}
&h(1-h)\frac{d^2R}{dh^2}+(1-D_{*}h)\frac{dR}{dh}+\nonumber \\
&\left[\frac{P^2}{A(r_H)^2 h(1-h)}-
\frac{\Lambda}{r_H^2 A(r_H)^2 (1-h)}\right]R=0\,,
\end{align}
where 
\begin{align}
&P=\omega(r_H+a^2/r_H)-ma/r_H\,,\\
&\Lambda=[l(l+n+1)+j(j+n-1)a^2/r_H^2](r_H^2+a^2)\,,\\ 
&D_{*}=1-\frac{4a^2 r_H^2}{\left[(n+1)r_H^2+(n-1)a^2\right]^2}\,.
\end{align}
Using the redefinition $R(h)=h^{\alpha}(1-h)^{\beta}F(h)$ the above equation takes the form
\be\label{NH1rot}
h(1-h)\frac{d^2F}{dh^2}+[c-(a+b+1)h]\frac{dF}{dh}-
(ab)F=0\,,
\ee
with 
\be
a=\alpha+\beta+D_{*}-1\,,\hspace{3mm} b=\alpha+\beta\,,\hspace{3mm} c=1+2\alpha\,,
\ee
and $\alpha$ and $\beta$ must satisfy the following algebraic equations  
\begin{eqnarray}
&& \alpha^2+\frac{P^2}{A(r_H)^2}=0\,,\\
&& \beta^2+\beta(D_{*}-2)+\frac{P^2}{A(r_H)^2}-\frac{\Lambda}{r_H^2 A(r_H)^2}=0\,,
\end{eqnarray}
whose solutions read
\begin{eqnarray}
 \alpha_{\pm}&=&\pm i\frac{P}{A(r_H)}\,,\\
\beta_{\pm}&=&\frac{1}{2}\Bigg[(2-D_{*})\nonumber\\
&&\pm\sqrt{(D_{*}-2)^2-4\frac{P^2}{A(r_H)^2}+4\frac{\Lambda}{r_H^2 A(r_H)^2}}\,\Bigg]\,.
\end{eqnarray}
The two linearly independent solutions of Eq.~\eqref{NH1rot} are $F(a,b;c;h)$ and $h^{1-c}F(a+1-c,b+1-c;2-c;h)$, where $F$ is the hypergeometric function. Convergence requires $\rm{Re}\left[c-a-b\right]>0$, which can be only obtained if the minus sign is chosen in the solutions above. In the following, we shall identify $\beta\equiv\beta_-$ and $\alpha\equiv\alpha_-$.
The general solution of Eq.~(\ref{NHrot}) is then
\begin{align}
&R(h)=A_1 h^{\alpha}(1-h)^{\beta}F(a,b,c;h)\nonumber\\
&+B_1 h^{-\alpha}(1-h)^{\beta}F(a+1+c,b+1-c,2-c;h).
\end{align}
Expanding the above result near the horizon, we get
\be
R(h)=A_1 h^{-i\frac{P}{A(r_H)}}+B_1 h^{i\frac{P}{A(r_H)}}=A_1 e^{-ik_H r_{*}}+B_1 e^{ik_H r_{*}}.
\ee
Regularity at the horizon requires $B_1=0$. The near-horizon solution can be written as (see~\cite{handmath})
\begin{align}
&R(h)=A_1 h^{\alpha}(1-h)^{\beta}
\frac{\Gamma[1+2\alpha]\Gamma[2-D_{*}-2\beta]}{\Gamma[2-D_{*}+\alpha-\beta]
\Gamma[1+\alpha-\beta]}\times\nonumber\\
&F(a,b,a+b-c;1-h)+A_1 h^{\alpha}(1-h)^{2-D_{*}-\beta}\times \nonumber\\
&\frac{\Gamma[1+2\alpha]\Gamma[2\beta+D_{*}-2]}{\Gamma[\alpha+\beta+D_{*}-1]
\Gamma[\alpha+\beta]}\times\nonumber\\
&F(c-a,c-b,c-a-b+1;1-h).
\end{align}
We can now expand this result in the low-frequency regime and for small values of $a/r_H$, in the region where $1-h\ll 1$ and $r\gg r_H$, 
\begin{align}\label{NH2rot}
&R\sim\frac{X_{kjm}^{r_H}}{r^{1+n/2}}\nonumber\\
&\sim\frac{r^l}{(2M)^{\frac{2l+n+1}{2(n+1)}}r_H^{\frac{1}{2}}}
\frac{\Gamma[1+2\alpha]\Gamma[2-D_{*}-2\beta]}{\Gamma[2-D_{*}+\alpha-\beta]
\Gamma[1+\alpha-\beta]}\,.
\end{align}

We will follow Poisson to solve the wave equation at large distances~\cite{poisson1}. It is useful to rewrite the radial equation~(\ref{radial}) in terms of the dimensionless variable $z=\omega r$. At large distances Eq.~\eqref{radial} reads
\begin{align}\label{approxrot}
&\Big[f\frac{d^2}{dz^2}+\frac{(n+1)\epsilon}{z^{2+n}}\frac{d}{dz}+1
-\frac{l(l+n+1)+\frac{n}{2}(1+\frac{n}{2})}{z^2}\nonumber\\
&-\frac{\epsilon\left[1+n\left(\frac{n}{4}+1\right)\right]}{z^{3+n}}\Big]X_{{kjm}}(z)=0\,,
\end{align}
where $f=1-\epsilon/z^{n+1}$, $\epsilon=2M \omega^{n+1}$ is a dimensionless parameter, and we used the fact that at large distances the eigenvalues take the form $A_{ljm}=l(l+n+1)$. We can rewrite it in a simpler form if we define the quantum number $J(J+1)=l(l+n+1)+\frac{n}{2}(1+\frac{n}{2})$. Solving for $J$, and assuming $J$ is a non-negative number, we get
\be \label{quantumJrot}
J=l+\frac{n}{2}.
\ee 
In the limit $\epsilon\ll 1$, Eq.~\eqref{approxrot} reads
\be\label{eqauxismallrot}
\left[\frac{d^2}{dz^2}+1-\frac{J(J+1)}{z^2}\right]X_{{kjm}}(z)=0\,.
\ee
The solution can be written in terms of a linear combination of Riccati-Bessel functions, $\sqrt{z}J_{J+1/2}(z)$ and $\sqrt{z}N_{J+1/2}(z)$.
The requirement that $X^{r_H}_{{kjm}}(z)$ be regular at the horizon demands
\be\label{Xhorizonrot}
X^{r_H}_{{kjm}}(z)=B \sqrt{z}J_{J+1/2}(z)\,,
\ee
where $B$ is a constant. The asymptotic expansions for the Bessel functions are well known and read 
\be \label{outerrot}
X^{r_H}_{{kjm}}(z \ll 1)\sim \frac{B z^{J+1}}{2^{J+1/2}\Gamma[J+3/2]}\left[1+O(z^2)\right]\,.
\ee
At large distance, the second term within the square brackets is subdominant and we shall ignore it.
Matching~(\ref{outerrot}) to~(\ref{NH2rot}) we get
\begin{align}
B&=\frac{2^{J+1/2}\Gamma[J+3/2]\Gamma[1+2\alpha]\Gamma[2-D_{*}-2\beta]}{\Gamma[2-D_{*}+\alpha-\beta]
\Gamma[1+\alpha-\beta]}\times \nonumber\\
&\frac{(2M)^{1/(2n+2)}}{\epsilon^{(J+1)/(n+1)}r_H^{1/2}}\left[1+O(\epsilon)\right]\,.
\end{align}
The parameter $A_{\rm in}$ can be extracted from the behavior of the function near $z=\infty$. Recalling the large-argument of the Bessel functions, $X^{r_H}_{{kjm}}(z\rightarrow\infty)\sim B\sqrt{2/\pi}\sin(z-J\pi/2)$ and, using Eq.~(\ref{bound}), it follows that 
\begin{align}
A_{\rm in}&=\frac{2^{J}\Gamma[J+3/2]\Gamma[1+2\alpha]\Gamma[2-D_{*}-2\beta]}{\sqrt{\pi}\Gamma[2-D_{*}+\alpha-\beta]
\Gamma[1+\alpha-\beta]}\times \nonumber\\
&\left(\frac{i}{\epsilon^{1/(n+1)}}\right)^{J+1}\frac{(2M)^{1/(2n+2)}}{r_H^{1/2}}
\left[1+O(\epsilon)\right]\,.
\end{align}

With all of this at hand, we can now compute the flux at infinity in the low-frequency regime. 
From Eqs.~(\ref{insol}) and~(\ref{flux}) we get
\begin{align}
&\dot{E}_{\infty}=m^2\Omega_p^2\left|Z^{\infty}_{{kjm}}\right|^2=\nonumber\\
&=m^{2+2l+n}
\left[\frac{\alpha m_p\sqrt{\pi}}{2^{l+n/2+1}\Gamma[l+n/2+3/2]}\right]^2\times\nonumber\\
&\left[(n+1)M\right]^{l+n/2+1}
|S_{kjm}(\pi/2)|^2|Y_{j}(\pi/2,\pi/2,\ldots)|^2\nonumber\\
&\times r_0^{-\frac{2l(n+1)+(n+2)(n+3)}{2}}\,.\label{anafluxinf}
\end{align}
where we used the fact that for small frequencies (large distances) $r^2+a^2\sim r^2$,
$U^t\sim 1$ and $\omega=m\Omega_p\sim m\sqrt{(n+1)M}r_0^{-(3+n)/2}$. 

Let us now perform the same calculation for the solution $X_{{kjm}}^{\infty}$, which satisfies outgoing-wave boundary conditions at infinity. The method is analogous to that already described above. However, since for this case the boundary condition is imposed at infinity we do not require regularity at the horizon. In the limit $\epsilon\ll 1$, $X_{{kjm}}^{\infty}$ can be identified, up to a normalization constant, with
\be
X^{\infty}_{{kjm}}(z)=C\,\sqrt{z}\,H_{J+1/2}^{(1)}(z)\,,
\ee
where $H_{J+1/2}^{(1)}(z)=J_{J+1/2}(z)+iN_{J+1/2}(z)$ is the Hankel function. To determine the constant $C$ we match this solution in the limit $z\rightarrow \infty$ to the required boundary condition
\be 
X^{\infty}_{{kjm}}(z\rightarrow \infty)\sim e^{iz}\,.
\ee
Recalling the asymptotic behavior of the Hankel functions, we have
\be
C \sqrt{z} H_{J+1/2}^{(1)}(z\rightarrow\infty)\sim C \sqrt{\frac{2}{\pi}}e^{iz}[(-i)^{J+1}+O(1/z)]\,,
\ee
from which we get $C=i^{J+1}\sqrt{\pi/2}$.

The small-argument behavior of the Bessel functions reads (when $J>0$)
\begin{align}
J_{J+1/2}(z\ll 1)\sim \frac{z^{J+1/2}}{2^{J+1/2}\Gamma[J+3/2]}\left[1+O(z^2)\right]\,,\nonumber\\
N_{J+1/2}(z\ll 1)\sim -\frac{2^{J+1/2}}{\pi}\frac{\Gamma[J+1/2]}{z^{J+1/2}}\left[1+O(z^2)\right]\,.
\end{align}
At leading order and near $z=0$, the function $N_{J+1/2}$ dominates over $J_{J+1/2}$. Hence, we get
\be 
X^{\infty}_{{kjm}}(z\ll 1)\sim i^J\frac{2^J}{\sqrt{\pi}}\Gamma[J+1/2]z^{-J}\,.
\ee

We can now compute the flux across the horizon. Using Eqs.~(\ref{horsol}) and~(\ref{flux}) we get 
\begin{align}
&\dot{E}_{H}=m \Omega_p k_H\left|Z^{r_H}_{{kjm}}\right|^2=\nonumber\\
&=m k_H (\alpha\,m_p)^2 \Gamma_1^2 \left(\frac{n+1}{2}\right)^{1/2} r_H(2M)^{\frac{2l+\frac{3n}{2}+\frac{3}{2}}{n+1}}\times \nonumber\\
&|S_{kjm}(\pi/2)|^2|Y_{j}(\pi/2,\pi/2,\ldots)|^2 \times r_0^{-\frac{4l+5n+7}{2}}\,,
\end{align}
where $\Gamma_1=\frac{\Gamma[l+n/2+1/2]\Gamma[2-D_{*}+\alpha-\beta]
\Gamma[1+\alpha-\beta]}{2\Gamma[l+n/2+3/2]
\Gamma[1+2\alpha]\Gamma[2-D_{*}-2\beta]}$.

We can now obtain an expression for the ratio of the fluxes on the horizon and at infinity for general $l$, $m$ and $n$. Using the expressions for $\dot{E}_H$ and $\dot{E}_\infty$ calculated above, we find 
\begin{align}
&\frac{\dot E_H}{\dot{E_{\infty}}}=\frac{k_H r_H\left[2^{l+n/2+1}\Gamma[l+n/2+3/2]\right]^2}{\pi m^{2l+n+1}}\Gamma^2_1\nonumber\\
&\times\left(\frac{2}{n+1}\right)^{\frac{2l+n+1}{2}}
(2M)^{-\frac{(n-1)(2l+n+1)}{2(1+n)}} r_0^{\frac{(n-1)(n+1+2l)}{2}}\,,
\end{align}
This can be written as a function of the orbital velocity,
\begin{eqnarray}\label{ratioflux}
\frac{\dot E_H}{\dot{E_{\infty}}}&=&\frac{k_H r_H \left[2^{l+n/2+1}\Gamma[l+n/2+3/2]\right]^2}{\pi m^{2l+n+1}}\Gamma^2_1\times\nonumber\\
&&\left(\frac{2}{n+1}\right)^{\frac{1+2l+n}{1+n}}
\times v^{-\frac{(n-1)(n+1+2l)}{n+1}}\,,
\end{eqnarray}
where we have used Eq.~\eqref{vel} at large distance.
For sufficiently small orbital frequencies, such that the superradiance condition is met, the flux at the horizon is negative, we then find that the ratio between the fluxes {\it grows} in magnitude with $r_0$ and the particle is tidally accelerated outwards.
For the dipolar mode, $l=1$, this expression is in complete agreement with the expected behavior derived from a newtonian tidal analysis, cf. Eq.~(\ref{tideratio}). 

\begin{figure}[htb]
\begin{center}
% \begin{tabular}{c}
\epsfig{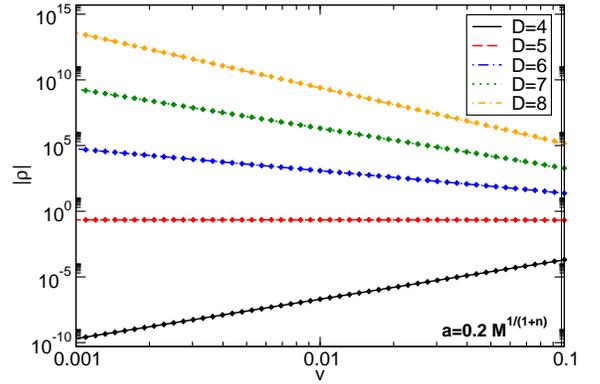}
% \end{tabular}
\caption{Comparison between the flux ratio $\rho=\dot E_H^{\rm Tot}/\dot E_{\infty}^{\rm Tot}$ (in absolute value) calculated analytically and numerically, as a function of the particle velocity $v$ for $n=0,1,2,3,4$ ($D=4,5,6,7,8$) and $a=0.2M^{1/(1+n)}$. The straight curves correspond to the analytical formula with $l=1$ and the dots are the numerical results discussed in Sec.~\ref{sec:numerical}. In the slowly-rotating regime, numerical results are in very good agreement with the analytical formula. \label{fig:flux_comp}}\end{center}
\end{figure}

Note that these results were derived under the assumption of slow rotation, $a\ll r_H$. This approximation is particularly severe in the near-extremal, five-dimensional case, where $r_H\to0$. Nevertheless, as we discuss in the next section, our method captures the correct scaling of the energy fluxes for \emph{any} spin and it even gives overall coefficients which are in very good agreement with the numerical ones in the slowly-rotating case. This is shown in Fig.~\ref{fig:flux_comp}, where we compare the analytical results of this section with the numerical fluxes computed in Sec.~\ref{sec:numerical}.
%

%%%%%%%%%%%%%%%%%%%%%%%%%%%%%%%%%%%%%%%%%%%%%%%%%%%%%%%%%%%%%%%%%%%%%%%%%%%%%
\section{Numerical Results}\label{sec:numerical}
%%%%%%%%%%%%%%%%%%%%%%%%%%%%%%%%%%%%%%%%%%%%%%%%%%%%%%%%%%%%%%%%%%%%%%%%%%%%%
The Green function approach described above can be implemented numerically using standard methods~\cite{Detweiler:1978ge,Cardoso:2011xi,Yunes:2011aa}. For given values of $r_0$, $a$ and $n$, we can compute the fluxes by truncating the sum in Eq.~\eqref{flux} to some $k_{\rm max}$, $m_{\rm max}$ and $j_{\rm max}$. As discussed before, for circular orbits only $j=0$ terms give a nonvanishing contribution. 

For small and moderately large orbital velocities, the sum converges rapidly even for small truncation orders and we typically set $k_{\rm max}=3$ and $m_{\rm max}=6$. 
However, the convergence is very poor when the orbital velocity approaches the speed of light, i.e. when the orbit is close to the prograde null circular geodesic. Recall that circular orbits around Myers-Perry BHs in higher dimensions are unstable~\cite{Cardoso:2008bp} and, in particular there is no innermost stable circular orbit for $D>4$. Thus, for our purposes we could in principle consider particles in circular orbit up to the light-ring, which exists for any dimension~\cite{Cardoso:2008bp}. As the particle approaches the light-ring, the flux is dominated by increasingly higher multipoles, thus affecting the convergence properties of the series~\eqref{flux}. For this reason, the plots presented below are extended up to some value of the velocity that guarantees good convergence.

Furthermore, for large orbital velocity and highly spinning BHs, the zeroth order angular eigenfunctions and the corresponding eigenfrequencies~\eqref{Akjm} might be not accurate. Therefore, when $a\omega\gtrsim1$, we have used exact numerical values of $A_{kjm}$ obtained by solving Eq.~\eqref{ang} with the continued fraction
method~\cite{Berti:2005gp}. We note however that Eq.~\eqref{Akjm} reproduces the exact results surprisingly well, even when $a\omega\sim1$.

We checked our method by reproducing the results of Ref.~\cite{Yunes:2011aa} for the massless case in four dimensions. In addition, we can compute the energy flux in any number of dimensions.
The fluxes $\dot E_H$ and $\dot E_\infty$ for $D=5$ and $D=6$ are shown in Tables~\ref{tab:flux1} and~\ref{tab:flux2} for $r_0=10 r_H$. We show the total flux as well as the first multipolar contributions.

%%%%%%%%%%%%%%%%%%%%%%%
 \begin{table}[th!]
%%%%%%%%%%%%%%%%%%%%%%%
\caption{\label{tab:flux1} Fluxes across the horizon and to infinity for $n=1$ ($D=5$), $a=M^{1/(1+n)}$ and $r_0=10 r_H$. In the last row we show the total flux obtained summing up to $k_{\rm max}=3$ and $m_{\rm max}=6$.}
%%%%%%%%%%%%%%%%%%%%%%
\begin{tabular}{c|c|c|c|c|c|c}
\hline
\hline
$k$ & $m$ & $j$ & $r_0/r_H$ & $\dot E_H(\alpha m_p)^{-2}$ & $\dot E_{\infty}(\alpha m_p)^{-2}$ &	 $|\dot E_H|/\dot E_{\infty}$\\
\hline 
$0$	&	$1$	& $0$ &	$10$	&	$-1.4759 \times 10^{-9}$ & $1.5790 \times 10^{-9}$ & $0.9347$\\
$0$	&	$2$	& $0$ &	$10$	&	$-8.3175 \times 10^{-11}$ & $1.6288 \times 10^{-10}$ & $0.5107$\\
$0$	&	$3$	& $0$ &	$10$	&	$-3.1691 \times 10^{-12}$ & $1.0080 \times 10^{-11}$ & $0.3144$\\
$1$	&	$1$	& $0$ &	$10$	&	$-1.2718 \times 10^{-14}$ & $5.1192 \times 10^{-16}$ & $24.844$\\
\hline
\multicolumn{3}{c|}{$\sum_{kmj}$}   &$10$ & $-3.1248\times 10^{-9}$ & $3.5052\times 10^{-9}$ & $0.8915$\\
\hline
\end{tabular}
\end{table}
%%%%%%%%%%%%%%%%%%%%%%%%%%%%%%%%%%%
 \begin{table}[th!]
%%%%%%%%%%%%%%%%%%%%%%%%%%%%%%%%%%%
\caption{\label{tab:flux2} Same as in Table~\ref{tab:flux1} but for $n=2$ ($D=6$).
}
%%%%%%%%%%%%%%%%%%%%%%%%%%%%%%%%%%
\begin{tabular}{c|c|c|c|c|c|c}
\hline
\hline
$k$ & $m$ & $j$ & $r_0/r_H$ & $\dot E_H(\alpha m_p)^{-2}$ & $\dot E_{\infty}(\alpha m_p)^{-2}$ &	 $|\dot E_H|/\dot E_{\infty}$\\
\hline
$0$	&	$1$	& $0$ &	$10$	&	$-4.1768 \times 10^{-12}$ & $1.3915 \times 10^{-14}$ & $300.155$\\
$0$	&	$2$	& $0$ &	$10$	&	$-3.0159 \times 10^{-13}$ & $3.8070 \times 10^{-16}$ & $792.180$\\
$0$	&	$3$	& $0$ &	$10$	&	$-1.3677 \times 10^{-14}$ & $4.8651 \times 10^{-18}$ & $2811.35$\\
$1$	&	$1$	& $0$ &	$10$	&	$-2.8126 \times 10^{-16}$ & $1.0497 \times 10^{-22}$ & $2.6796\times 10^6$\\
\hline
\multicolumn{3}{c|}{$\sum_{kmj}$}   &$10$ & $-8.9858\times 10^{-12}$ & $2.8601\times 10^{-14}$ & $314.176$\\
\hline
\end{tabular}
\end{table}
%%%%%%%%%%%%%%%%%%%%%%%%%%%%%%%%%%%%%%%%%%%

\begin{figure}[htb]
\begin{center}
% \begin{tabular}{c}
\epsfig{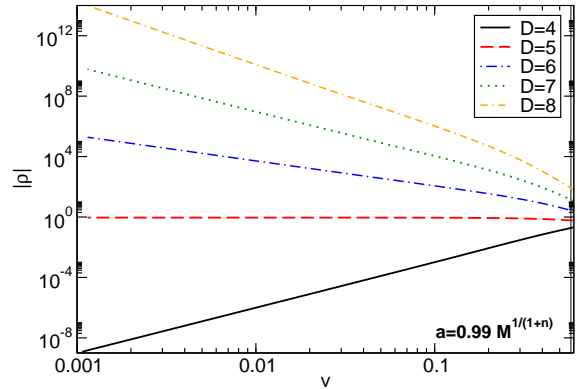}
% \end{tabular}
\caption{The flux ratio $\rho=\dot E_H^{\rm Tot}/\dot E_{\infty}^{\rm Tot}$ (in absolute value) as a function of the particle velocity $v$ defined in Eq.~\eqref{vel} for $n=0,1,2,3,4$ ($D=4,5,6,7,8$) and $a=0.99M^{1/(1+n)}$. \label{fig:flux_n}}
\end{center}
\end{figure}

\begin{figure*}[htb]
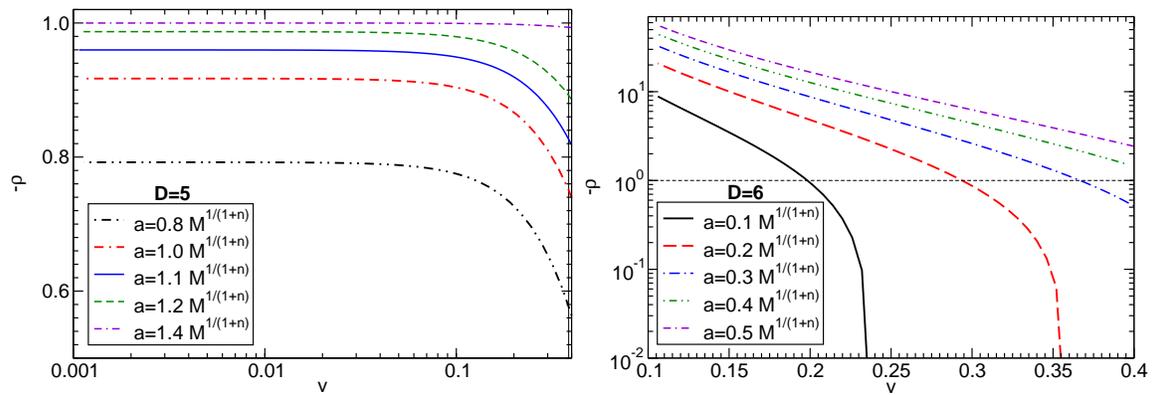

\begin{center}
\begin{tabular}{cc}
\epsfig{file=fluxes_n1.eps,width=7.5cm,angle=0,clip=true}&
\epsfig{file=fluxes_n2.eps,width=7.5cm,angle=0,clip=true}
\end{tabular}
\caption{The ratio $\rho=\dot E_H^{\rm Tot}/\dot E_{\infty}^{\rm Tot}$ as a function of the orbital velocity defined in Eq.~\eqref{vel} for several values of $a$. Left panel: when $D=5$, the ratio is constant in the small $v$ region and it approaches unity in the extremal limit, $a\to\sqrt{2M}$. Right panel: when $D=6$, the flux at the horizon can exceed the flux at infinity. For each curve, the intersection with the horizontal line corresponds to a floating orbit, $-\rho=1$. Note that, at large orbital velocity, the superradiant condition is not met and $\dot{E}_H>0$. \label{fig:flux5d6d}}
\end{center}
\end{figure*}

Tables~\ref{tab:flux1} and~\ref{tab:flux2} confirm our analytical expectations that the behavior for $n>1$ ($D>5$) is qualitatively different: the energy flux across the horizon is larger (in modulus) than the flux at infinity. This is shown in Fig.~\ref{fig:flux_n}, where we compare the flux ratio $\rho=\dot E^{\rm Tot}_H/\dot E^{\rm Tot}_{\infty}$ as a function of the orbital velocity $v$ for $a=0.99M^{1/(1+n)}$ in various dimensions. 
This figure is analogous to Fig.~\ref{fig:flux_comp} but for $a=0.99M^{1/(1+n)}$, i.e. a regime that is not well described by the analytical formula~\eqref{ratioflux}.
For $D=4$, we find the usual behavior, i.e. the flux at the horizon is usually negligible with respect to that at infinity and the ratio decreases rapidly at large distance. The case $D=5$ marks a transition, because $\rho$ is constant at large distance. This is better shown in the left panel of Fig.~\ref{fig:flux5d6d}. On the other hand, for any $D>5$ the flux across the horizon generically dominates over the flux at infinity.
%

% %
% \begin{figure}[htb]
% \begin{center}
% % \begin{tabular}{c}
% \epsfig{file=Plots/circ_null.eps,width=7.5cm,angle=0,clip=true}
% % \end{tabular}
% \caption{Radius of the prograde null circular orbit as a function of the rotation parameter in various dimensions. For $D=4,5$ curves terminate at extremality.\label{fig:circ_null}}
% \end{center}
% \end{figure}
% %

In Fig.~\ref{fig:flux5d6d} we show the flux ratio $\rho$ for some selected value of the spin parameter $a$ in five dimensions (left panel) and in six dimensions (right panel). When $D=5$, the ratio is constant in the small $v$ region and it approaches unity in the extremal limit, $a\to\sqrt{2M}$. As shown in the right panel of Fig.~\ref{fig:flux5d6d}, when $D=6$ there exist some orbital velocity for which $-\rho=1$, corresponding to a total vanishing flux, $\dot{E}_H+\dot{E}_\infty=0$. These orbital frequencies correspond to ``floating'' orbits~\cite{teunature,Cardoso:2011xi}.
Although in the right panel of Fig.~\ref{fig:flux5d6d} this is shown only for $a/M^{1/3}=0.1,0.2,0.3$, we expect this to be a generic feature also for larger values of the spin. The poor convergence properties of the series~\eqref{flux} prevent us to extend the curves to larger values of $v$, where floating orbits for $a>0.3M^{1/3}$ are expected to occur. 
At smaller velocity, the energy flux contribution dominates and the motion of the test-particle is generically dominated by tidal acceleration. Similar results can be obtained for any $D\geq6$.

%%%%%%%%%%%%%%%%

%%%%%%%%%%%%%%%%%%%%%%%%%%%%%%%%%%%%%%%%%%%%%%%%%%%%%%%%%%%%%%%%%%%%%%%%%%%%%%
\section{Conclusions}\label{sec:conclusion}
%%%%%%%%%%%%%%%%%%%%%%%%%%%%%%%%%%%%%%%%%%%%%%%%%%%%%%%%%%%%%%%%%%%%%%%%%%%%%%
We computed the rate at which the energy is extracted from a singly-spinning, higher-dimensional BH when a massless scalar field is coupled to a test-particle in circular orbit. We showed that, for dimensions greater than five and small orbital velocities, the energy flux radiated to infinity becomes negligible compared to the energy extracted from the BH via superradiance. 

Although we considered scalar-wave emission, we expect our results to be generic in higher dimensions. In particular, superradiance should be a dominant effect also for gravitational radiation. At leading order, the ratio $|\dot E_H|/\dot E_{\infty}$ for gravitational radiation should scale with the velocity as described by Eq.~(\ref{ratioflux}). The dominant quadrupole term ($l=2$) reads~\cite{Cardoso:2012zn}
%%%
\begin{equation}
 \frac{|\dot E_H|}{\dot E_{\infty}}\sim v^{-\frac{(n-1)(n+5)}{n+1}}\,.\nonumber
\end{equation}
%%%
By comparing the formula above to Eq.~\eqref{ratioflux} with $l=1$, we note that dipolar effects are dominant over their quadrupolar counterpart. Nevertheless, even in the purely gravitational case, tidal acceleration and floating orbits around spinning BHs are generic and distinctive effects of higher dimensions.

In principle, gravitational waveforms would carry a clear signature of floating orbits \cite{Cardoso:2011xi,Yunes:2011aa}. Does floating or these strong tidal effects 
have any significance in higher-dimensional BH physics? We should start by stressing that circular geodesics in higher dimensions are unstable,
on a timescale comparable to the one discussed here~\cite{Cardoso:2008bp}; however, our analysis suggests that, while more pronounced for circular orbits, tidal
acceleration is generic and in no way dependent on the stability of the orbit under consideration. We are thus led to conjecture that tidal effects are crucial to determine binary evolution in higher dimensions. It is possible that tidal effects already play a role in the numerical simulations of the kind recently reported in Refs.~\cite{Witek:2010xi,Okawa:2011fv,Cardoso:2012qm}, but further study is necessary. One of the consequences of our results for those type of simulations is, for instance, that in higher dimensional BH collisions the amount of gravitational radiation accretion might play an important role. It would certainly be an interesting topic for further study to understand tidal effects for generic orbits, and to include finite-size effects in the calculations.

%%%%%%%%%%%%%%%%%%%%%%%%%%%%%%%%%%%%%%%%%%%%%%%%%%%%%%%%%%%%%%%%%%%%%%%%%%%%%%%%%%%%%%
\begin{acknowledgments}
  This work was supported by the DyBHo--256667 ERC Starting Grant, the
  NRHEP--295189 FP7--PEOPLE--2011--IRSES Grant, and by FCT - Portugal through PTDC
  projects FIS/098025/2008, FIS/098032/2008, CTE-ST/098034/2008,
  CERN/FP/123593/2011. P.P. acknowledges financial support provided by
  the European Community through the Intra-European Marie Curie contract
  aStronGR-2011-298297 and the kind hospitality of the International School for Advanced Studies (SISSA) in Trieste, during the last stages of this work.
\end{acknowledgments}
%%%%%%%%%%%%%%%%%%%%%%%%%%%%%%%%%%%%%%%%%%%%%%%%%%%%%%%%%%%%%%%%%%%%%%%%%%%%%%%%%%%%%%%
\bibliography{ref}
\end{document}